\documentclass[longauth]{aa}

\usepackage{longtable}    
\usepackage{latexsym}
\usepackage{graphicx}
\usepackage{epsfig}
\usepackage{lscape}
\usepackage[authoryear]{natbib}
\bibpunct{(}{)}{;}{a}{}{,} % to follow the A&A style

\def \ms{m\,s$^{-1}$\/}

\begin{document}

\newcommand{\msun}{\mbox{M$_\odot$}}
\newcommand{\rsun}{\mbox{R$_\odot$}}
\newcommand{\lsun}{\mbox{L$_\odot$}}
\newcommand{\mjup}{\mbox{M$_{\rm Jup}$}}
\newcommand{\rjup}{\mbox{R$_{\rm Jup}$}}
\newcommand{\mear}{\mbox{M$_\oplus$}}
\newcommand{\rear}{\mbox{R$_\oplus$}}

\newcommand{\sqdeg}{\mbox{deg$^{2}$}}
\newcommand{\mini}{\mbox{$M_{\rm i}$}}
\newcommand{\mto}{\mbox{$M_{\rm TO}$}}
\newcommand{\thef}{\mbox{$t_{\rm HeF}$}}
\newcommand{\Mhef}{\mbox{$M_{\rm HeF}$}}
\newcommand{\diff}{\mbox{${\rm d}$}}
\newcommand{\sub}[1]{\mbox{$_{\rm #1}$}}
\newcommand{\bvo}{\mbox{$(B\!-\!V)_{0}$}}
\newcommand{\ubo}{\mbox{$(U\!-\!B)_{0}$}}
\newcommand{\ub}{\mbox{$U\!-\!B$}}
\newcommand{\bv}{\mbox{$B\!-\!V$}}
\newcommand{\vi}{\mbox{$V\!-\!I$}}
\newcommand{\vr}{\mbox{$V\!-\!R$}}
\newcommand{\vk}{\mbox{$V\!-\!K$}}
\newcommand{\jh}{\mbox{$J\!-\!H$}}
\newcommand{\jk}{\mbox{$J\!-\!K$}}
\newcommand{\yk}{\mbox{$Y\!-\!K$}}
\newcommand{\jks}{\mbox{$J\!-\!K_{\rm s}$}}
\newcommand{\yks}{\mbox{$Y\!-\!K_{\rm s}$}}
\newcommand{\ks}{\mbox{$K_{\rm s}$}}
\newcommand{\hk}{\mbox{$H\!-\!K$}}
\newcommand{\mv}{\mbox{$M_{V}$}}
\newcommand{\mi}{\mbox{$M_{I}$}}
\newcommand{\mk}{\mbox{$M_{K}$}}
\newcommand{\mbol}{\mbox{$m_{\rm bol}$}}
\newcommand{\Mbol}{\mbox{$M_{\rm bol}$}}
\newcommand{\dm}{\mbox{$(m\!-\!M)$}}
\newcommand{\dmo}{\mbox{$(m\!-\!M)_{0}$}}
\newcommand{\fb}{\mbox{$f_{\rm b}$}}
\newcommand{\av}{\mbox{$A_V$}}
\newcommand{\ebv}{\mbox{$E_{B\!-\!V}$}}
\newcommand{\eyj}{\mbox{$E_{Y\!-\!J}$}}
\newcommand{\eyk}{\mbox{$E_{Y\!-\!K}$}}
\newcommand{\feh}{\mbox{\rm [{\rm Fe}/{\rm H}]}}
\newcommand{\mh}{\mbox{\rm [{\rm M}/{\rm H}]}}
\newcommand{\Teff}{\mbox{$T_{\rm eff}$}}
\newcommand{\logT}{\mbox{$\log\Teff$}}
\newcommand{\logL}{\mbox{$\log L/L_{\odot}$}}
\newcommand{\logg}{\mbox{$\log g$}}
\newcommand{\co}{\mbox{${\rm C/O}$}}
\newcommand{\chisqmin}{\mbox{$\chi^2_{\rm min}$}}
\newcommand{\etal}{\mbox{\rm et~al.}}
\newcommand{\hii}{\mbox{H\,{\sc ii}}}
\newcommand{\comment}[1]{}
\newcommand{\beq}{\begin{equation}}
\newcommand{\eeq}{\end{equation}}
\newcommand{\beqa}{\begin{eqnarray}}
\newcommand{\eeqa}{\end{eqnarray}}
\newcommand{\abs}[1]{\mid #1 \mid}
\newcommand{\norm}[1]{\| #1 \|}
\newcommand{\norms}[2]{\| #1 \cdot #2 \|}
\newcommand{\pscal}[2]{#1 \cdot #2}
\newcommand{\pvett}[2]{#1 \times #2}
\newcommand{\mineq}{\stackrel{<}{\sim}}
\newcommand{\maxeq}{\stackrel{>}{\sim}}
%funzioni matematiche
\newcommand{\sh}{\mathop{\rm sh}\nolimits}
\newcommand{\ch}{\mathop{\rm ch}\nolimits}
\newcommand{\setthsh}{\mathop{\rm settsh}\nolimits}
\newcommand{\settch}{\mathop{\rm settch}\nolimits}
\newcommand{\asin}{\mathop{\rm asin}\nolimits}
\newcommand{\acos}{\mathop{\rm acos}\nolimits}
\newcommand{\atan}{\mathop{\rm atan}\nolimits}
\newcommand{\tg}{\mathop{\rm tg}\nolimits}
\newcommand{\ctg}{\mathop{\rm ctg}\nolimits}
\newcommand{\ctgh}{\mathop{\rm ctgh}\nolimits}
\newcommand{\tgh}{\mathop{\rm tgh}\nolimits}
\newcommand{\atg}{\mathop{\rm atg}\nolimits}
\newcommand{\atgh}{\mathop{\rm atgh}\nolimits}
\newcommand{\actg}{\mathop{\rm actg}\nolimits}
\newcommand{\actgh}{\mathop{\rm actgh}\nolimits}
%calcolo differenziale
\newcommand{\Div}{\mathop{\rm div}\nolimits}
\newcommand{\rot}{\mathop{\rm rot}\nolimits}
\newcommand{\grad}{\mathop{\rm grad}\nolimits}
\newcommand{\de}{{\rm\,d}}
\newcommand{\integ}[2]{\int {#1}{\rm\,d} #2}
\newcommand{\intee}[4]{\int^{#1}_{#2} {#3}{\rm\,d} #4}
\newcommand{\dpar}[2]{\frac {\partial #1}{\partial #2}}
\newcommand{\dpars}[3]{\frac{\partial^#1 #2}{\partial #3^#1}}
\newcommand{\dparlog}[2]{\frac {\partial\,\ln #1}{\partial\,\ln #2}}
\newcommand{\deri}[2]{\frac {{\rm d} #1}{{\rm d} #2}}
\newcommand{\deris}[3]{\frac{{\rm d}^#1 #2}{{\rm d} #3^#1}}
\newcommand{\largeint}[3]{\begingroup\setbox0\hbox{$\displaystyle #3$}\setbox0\hbox{$\vcenter{\box0}$}\dimen0=\ht0 \advance\dimen0 by 2pt \ht0=\dimen0\dimen0=\dp0 \advance\dimen0 by 2pt \dp0=\dimen0\left\lmoustache{\vrule height \ht0 depth \dp0 width 0pt}_{\kern-0.5ex #1}^{#2}\box0\rigth.\endgroup}

\title{The GAPS Programme with HARPS-N at TNG\thanks{Based on observations collected at the Italian {\em Telescopio Nazionale Galileo} (TNG), 
operated on the island of La Palma by the Fundaci\'on Galileo Galilei of the INAF (Istituto Nazionale di Astrofisica) at the Spanish 
Observatorio del Roque de los Muchachos of the Instituto de Astrof\'isica de Canarias, in the frame of the programme 
{\em Global Architecture of Planetary Systems} (GAPS).}
}
\subtitle{I: Observations of the Rossiter-McLaughlin effect and characterisation of the transiting system 
Qatar-1\thanks{Also based on observations collected at Asiago Observatory, and Calar Alto Observatory.
Table\,\ref{tab:phot_data} is only available in electronic form at the CDS via anonymous ftp to 
cdsarc.u-strasbg.fr (130.79.128.5) or via http://cdsweb.u-strasbg.fr/cgi-bin/qcat?J/A+A/}
}

\author{
        E. Covino\inst{1} \and M. Esposito\inst{2,19} \and M. Barbieri\inst{3} 
\and L. Mancini\inst{4} \and V. Nascimbeni\inst{3,5}
\and R. Claudi\inst{5} \and S. Desidera\inst{5} \and R. Gratton\inst{5}
\and A. F. Lanza\inst{6}
\and A. Sozzetti\inst{7} \and K. Biazzo\inst{1,6}
\and L. Affer\inst{8}  
\and D. Gandolfi\inst{9} \and U. Munari\inst{5} \and I. Pagano\inst{6} \and A. S. Bonomo\inst{7} 
\and A. Collier Cameron\inst{10} 
\and G. H\'ebrard \inst{21} 
\and A. Maggio\inst{8} \and S. Messina\inst{6} \and  G. Micela\inst{8} 
\and E. Molinari\inst{11,12} 
\and F. Pepe\inst{13} \and G. Piotto\inst{3,5}
\and I. Ribas\inst{14} 
\and N. C. Santos\inst{15,16} 
\and J. Southworth\inst{17}
\and E. Shkolnik\inst{23} 
\and A.  H.M.J. Triaud\inst{13,24}
\and L. Bedin\inst{5} 
\and S. Benatti \inst{5} \and C. Boccato\inst{5} \and  M. Bonavita\inst{5} \and F. Borsa\inst{20,26} \and  L. Borsato\inst{3} 
\and D. Brown\inst{10} 
\and E. Carolo\inst{3}  \and S. Ciceri\inst{4}
\and R. Cosentino\inst{11} 
\and M. Damasso\inst{3,27,7} 
\and F. Faedi\inst{18}
\and A. F. Mart\'{i}nez Fiorenzano\inst{11}
\and D.W. Latham\inst{25} \and C. Lovis\inst{13} \and C. Mordasini\inst{4} 
\and N. Nikolov\inst{4}
\and E. Poretti\inst{20} \and  M. Rainer\inst{20} 
\and R. Rebolo L\'opez\inst{2,19}
\and G. Scandariato\inst{6}
\and R. Silvotti\inst{7} 
\and R. Smareglia\inst{22} 
\and  J.~M. Alcal\'a\inst{1} 
\and  A. Cunial\inst{3} 
\and  L. Di Fabrizio\inst{2} 
\and  M.P. Di Mauro\inst{28} 
\and  P. Giacobbe\inst{7,27} 
\and  V. Granata\inst{3}
\and  A. Harutyunyan\inst{2}
\and  C. Knapic\inst{22}
\and  M. Lattanzi\inst{7} 
\and  G. Leto\inst{6}  
\and  G. Lodato\inst{29}
\and  L. Malavolta\inst{3}
\and  F. Marzari\inst{3}
\and  M. Molinaro\inst{22}
\and  D. Nardiello\inst{3}
\and  M. Pedani\inst{2}
\and  L. Prisinzano\inst{8}
\and  D. Turrini\inst{28}
}

\offprints{E. Covino}
\mail{covino@oacn.inaf.it}

\institute{INAF -- Osservatorio Astronomico di Capodimonte, via Moiariello, 16, 80131 Naples, Italy     % 1
 \and Instituto de Astrof\'isica de Canarias, C/ V\'ia L\'actea, s/n, E38205 - La Laguna (Tenerife), Spain    % 2
 \and Dipartimento di Fisica e Astronomia Galileo Galilei, Universit\`a di Padova,   % 3
          Vicolo dell'Osservatorio 2, I-35122, Padova, Italy
 \and Max-Planck-Institut f\"ur Astronomie, K\"onigstuhl 17, D-69117 Heidelberg, Germany  % 4
 \and INAF -- Osservatorio Astronomico di Padova, Vicolo dell'Osservatorio 5,  35122 Padova, Italy      % 5
 \and INAF -- Osservatorio Astrofisico di Catania, via S. Sofia 78, 95123 Catania, Italy     % 6
 \and INAF -- Osservatorio Astrofisico di Torino, Via Osservatorio 20, I-10025, Pino Torinese, Italy      % 7
 \and INAF -- Osservatorio Astronomico di Palermo, Piazza del Parlamento, Italy 1, I-90134, Palermo, Italy      % 8
 \and Research and Scientific Support Department, ESTEC/ESA, PO Box 299 2200 AG Noordwijk, The Netherlands      % 9
 \and  SUPA, School of Physics and Astronomy, University of St. Andrews, North Haugh, Fife KY16 9SS, UK     % 10
 \and  Fundaci\'on Galileo Galilei - INAF, Rambla Jos\'e Ana Fernandez P\'erez, 7 38712 Bre\~na Baja, TF - Spain     % 11
 \and  INAF -- IASF Milano, via Bassini 15, I-20133 Milano, Italy     % 12
 \and  Observatoire Astronomique de l'Universit´e de Geneve, 51 ch. des Maillettes - Sauverny, 1290, Versoix, Switzerland  %  13
 \and  Institut de Ci\`encies de l'Espai (CSIC-IEEC), Campus UAB, Facultat de Ci\`encies,        % 14
        Torre C5 parell, 2a pl, 08193 Bellaterra, Spain    % 14 %%%% 
 \and  Centro de Astrof\'isica, Universidade do Porto, Rua das Estrelas, 4150-762 Porto, Portugal    % 15
 \and  Departamento de F\'isica e Astronomia, Faculdade de Ci\^encias, Universidade do Porto, Portugal    % 16
 \and Astrophysics Group, Keele University, Staffordshire, ST5 5BG, UK     %  17 
 \and  Department of Physics, University of Warwick, Gibbet Hill Road, Coventry CV4 7AL, UK     % 18  
 \and  Departamento de Astrof\'isica, Universidad de La Laguna, Avda. Astrof\'isico Francisco  S\'anchez, s/n 38206 % 19
        La Laguna, Tenerife, Spain
 \and INAF -- Osservatorio Astronomico di Brera, Via E. Bianchi 46, 23807, Merate (LC), Italy  % 20
 \and Institut d'Astrophysique de Paris, UMR7095 CNRS, Universit\'e Pierre \& Marie Curie,  % 21
       98bis boulevard Arago, 75014 Paris, France 
 \and INAF -- Osservatorio Astronomico di Trieste, Via Tiepolo 11,  34143 Trieste Italy            % 22
 \and  Lowell Observatory, 1400 W. Mars Hill Road, Flagstaff, AZ, USA 86001    % 23
 \and Kavli Institute for Astrophysics and Space Research, Massachusetts Institute of Technology, Cambridge, MA 02139, USA % 24
 \and Harvard-Smithsonian Center for Astrophysics, 60 Garden Street, Cambridge, MA 02138, USA   % 25
 \and Universit\`a dell'Insubria - Dipartimento di Scienza e Alta Tecnologia, Via Valleggio 11, 22100 Como, Italy  % 26
  \and Osservatorio Astronomico della Regione Autonoma Valle d'Aosta, Fraz. Lignan 39, 11020, Nus (Aosta), Italy  % 27
  \and INAF Istituto di Fisica e Planetologia Spaziali, via Fosso del Cavaliere 100, 00133, Roma Italy % 28
  \and Dipartimento di Fisica dell'Universit\'a degli studi di Milano, via Celoria, 16, 20133 Milano Italy % 29
 }

\date{Received / accepted }

\abstract
% Context:
{Our understanding of the formation and evolution of planetary systems is still fragmentary 
becuase most of the current data provide limited information about the orbital structure and dynamics
of these systems.
The knowledge of the orbital properties for a variety of systems and at different ages
yields information on planet migration and on star-planet tidal interaction mechanisms.
}
% Aims:  
{In this context, a long-term, multi-purpose, observational programme has started with HARPS-N at TNG  
and aims to characterise the global architectural properties of exoplanetary systems.
The goal of this first paper is to fully characterise the orbital properties of the transiting system Qatar-1
as well as the physical properties of the star and the planet.
}
% Methods: 
{We exploit HARPS-N high-precision radial velocity measurements obtained during a transit
to measure the Rossiter-McLaughlin effect in the Qatar-1 system, and 
out-of-transit measurements to redetermine the spectroscopic orbit.
New photometric-transit light-curves were analysed and a spectroscopic characterisation of the host star 
atmospheric parameters was performed based on various methods (line equivalent width ratios, spectral 
synthesis,  spectral energy distribution).
} 
% Results:
{We achieved a significant improvement in the accuracy of the orbital parameters and derived
the spin-orbit alignment of the  system; this information, combined with the spectroscopic determination
of the host star properties (rotation, $T_{\rm eff}$, $\log{g}$, metallicity), allows us to derive the
fundamental physical parameters for star and planet (masses and radii).
The orbital solution for the  Qatar-1 system is consistent with a circular orbit  
and the system presents a sky-projected obliquity of $\lambda = -8.4 \pm 7.1$\,deg. 
The planet, with a mass of $1.33\pm0.05$\,M$_{\rm J}$, is found to be significantly more massive
than previously reported.
The host star is confirmed to be metal-rich ([Fe/H]= 0.20$\pm$0.10) and slowly rotating
(v$\sin{I} = 1.7 \pm 0.3 $\,km\,s$^{-1}$), though moderately active, as indicated by % the strength of 
the strong chromospheric emission in the \ion{Ca}{II} H\&K line cores ($\log{\rm R'_{HK}}\approx-4.60$).
}
% Conclusions: 
{We find that the system is well aligned and fits well within the general $\lambda$ versus $T_{\rm eff}$ trend.
We can definitely rule out any significant orbital eccentricity.
The evolutionary status of the system is inferred based on gyrochronology, and the present orbital 
configuration and timescale for orbital decay are discussed in terms of star-planet tidal interactions. 
}

	   
\titlerunning{GAPS. I. RM effect in Qatar-1}
\authorrunning{Covino et al.}
\maketitle

\section{Introduction}
\label{Sec:intro}
The study of extrasolar planets and the determination of their observational properties have made 
remarkable progress over the past decade. The surveys conducted so far with the most 
successful techniques (i.e. radial veocities and planetary transits photometry) have revealed planets 
spanning an unexpectedly wide range of orbital properties. Particularly surprising is the detection of 
close-in giant planets with orbital periods as short as one day, the so-called hot Jupiters.

Today ground-based and space-borne photometric surveys are yielding crucial information on 
transiting planets \citep{2013ApJS..204...24B}. However, the radial velocity technique is still not only 
of essential value (e.g., to confirm transiting planet candidates and determine their actual masses), 
but is now beginning to extend searches to still unexplored ranges of planet and host star properties, 
thanks to the improved sensitivity and stability of the new generation of high-resolution spectrographs, 
pioneered by the HARPS instrument on the 3.6m ESO telescope, and now followed by the newly built 
high-resolution spectrograph HARPS-N, recently come into operation on 
the Telescopio Nazionale Galileo, TNG \citep{2012SPIE.8446E..1VC}.

One of the relevant open questions in the exoplanetary field concerns the characterisation of the architectural 
properties of extrasolar planets and their possible dependence on the physical properties of the parent stars.
For example, known systems with giant planets are usually not followed-up to search for additional low-mass 
companions that might exist at a range of separations: as a consequence, we still lack information about the
frequency of solar-system-like systems. 
Properties such as frequency and orbital characteristics of exoplanets are expected to depend upon 
stellar properties, such as metallicity \citep{2009ApJ...697..544S,2011A&A...526A.112S,2013A&A...551A.112M}
and mass \citep{2010PASP..122..905J,2013A&A...549A.109B}, as well as on the host star's environment 
\citep{2007A&A...462..345D,2012A&A...545A.139P,2012ApJ...756L..33Q,2012A&A...542A..92R}.
Furthermore, the relative role of different mechanisms of the time evolution of planetary system architecture 
is still rather uncertain.

Crucial information on the mechanisms governing the evolution of planetary systems can be obtained 
by determining the relative orientation of the host-star spin axis, which is usually regarded as a relic 
of the angular momentum of the protostellar accretion disc, and the orbital axes of the planets.
Of the two main mechanisms invoked to explain inward  migration of giant planets 
from their original location, disc-planet interactions tend to preserve the initial spin-orbit 
alignment as well as orbit circularity, whereas dynamical interactions, such as planet-planet 
scattering \citep{2006RPPh...69..119P,2008ApJ...686..580C} or Kozai resonance due to 
the presence of an off-plane massive perturber \citep{2007ApJ...669.1298F}, are expected 
to alter both the inclination and eccentricity of the orbit.
Planet-planet scattering must occur after the disc dissipated, otherwise the disc interaction 
would drag the planet back on the median plane \citep{2011A&A...530A..41B,2009ApJ...705.1575M}. 
Although, as suggested recently by \citet{2013MNRAS.428..658T}, 
for a sufficiently high initial inclination of the orbital plane, even a disc could lead to Kozai cycles, 
and hence to spin-orbit misalignment.
The orbital eccentricity can become very high during Kozai cycles with consequent decrease of 
periastron distance and set-in of star-planet tidal interactions until the orbit eventually 
becomes very tight and circularised \citep{2011PhRvL.107r1101K,2011ApJ...742...94L}.

Determining the orientation of a planet orbital axis with respect to the stellar spin axis is 
therefore a way to assess the relative importance of the two migration mechanisms.
This can be accomplished through the observation of the Rossiter-McLaughlin (RM) effect, 
well known from the study of eclipsing binaries \citep{1924ApJ....60...15R,1924ApJ....60...22M}.
The RM effect is an anomaly in the radial velocity (RV) curve occurring during a planetary transit, 
whose shape yields information on the sky-projected angle $\lambda$ between the star spin axis and the 
planet orbital axis.
Therefore, measuring the RM effect provides a unique observational constraint 
to the actual spin-orbit misalignment \citep{2010ApJ...718L.145W}.
In some cases, the analysis of starspots can also provide a good determination of the sky-projected 
obliquity $\lambda$ and, in the most favourable ones, can even yield the true obliquity ($\psi$) of 
the orbit \citep{2011ApJ...740L..10N,2013MNRAS.428.3671T}.

So far, $\lambda$ has been measured for a growing number of transiting planets\footnote{
see~the~RM~encyclopedia: \\
http://www.aip.de/People/RHeller

and the TEPCat url: \\
http://www.astro.keele.ac.uk/jkt/tepcat/rossiter.html}
(over 60 to date), 
the majority of which do show values of $\lambda$ close to zero, pretty much like the planetary bodies 
orbiting our Sun, although a considerable fraction (nearly 40\%) shows substantial misalignment 
\citep{2012ApJ...757...18A}. 

In this paper and in a companion Letter (Paper\,II of the series) by Desidera et al. (2013), 
we present the first results obtained in the framework of the project {\it
Global Architecture of Planetary Systems} (GAPS), a large observational programme 
with HARPS-N which has recently been competitively awarded long-term status at the TNG.
GAPS is a structured, largely synergetic observational programme specifically designed 
to maximise the scientific return in several aspects of exoplanetary astrophysics, 
taking advantage of the unique capabilities provided by HARPS-N. The GAPS programme 
is composed of three main elements, including a) radial-velocity searches for low-mass 
planets around stars with and without known planets over a broad range of properties (mass, 
metallicity) of the hosts, b) characterisation measurements of known transiting systems, 
and c) improved determinations of relevant physical parameters (masses, radii, ages) 
and of the degree of star-planet interactions for selected planet hosts. 

In particular, within the framework of the GAPS programme element devoted to investigating 
the outcome of planet-disc and planet-planet interaction scenarios in exoplanet systems, 
we present here the first determination of the RM effect for the recently discovered transiting 
system \object{Qatar-1} ($\equiv$ GSC 04240-00470) \citep{2011MNRAS.417..709A}.
This system contains a hot Jupiter orbiting a V=12.84, metal-rich K-dwarf star, one of the faintest 
around which a planet has been discovered so far by ground-based surveys.
% Furthermore, as for other systems containing very hot Jupiters orbiting low-mass stars, Qatar-1 represents 
Furthermore, Qatar-1 represents an interesting study case for investigating the star-planet interaction and 
will set constraints on theories of tidal evolution for other systems that contain very hot Jupiters orbiting 
low-mass stars. 
For this purpose, it is mandatory to rely on refined orbital parameters as well as on accurate determinations 
of the physical properties for the star and the planet.

The plan of the paper is the following: in Sec.\,\ref{Sec:harps-n} we present the HARPS-N observations 
and in Sec.\,\ref{sec:phot-obs} we introduce the complementary photometric observations.
In Sec.\,\ref{Sec:analysis} we describe the analysis and present the results of the photometric and 
spectroscopic data, in Sec.\,\ref{sec:host_char} we derive the atmospheric properties of the host star 
based on different methods, while in Sec.\,\ref{Sec:results} we infer the system properties related to 
rotation and activity indicators and discuss them in terms of star-planet tidal interaction. 
In Sec.\,\ref{sec:conclusions} we summarise our results and main conclusions.

\section{HARPS-N observations and data reduction} 
\label{Sec:harps-n}
The spectroscopic observations of the transit were obtained on 2012 September 3, 
using the HARPS-N (High Accuracy Radial velocity Planet Searcher-North) spectrograph 
at the 3.58m Telescopio Nazionale Galileo (TNG) \citep{2012SPIE.8446E..1VC}.
HARPS-N, a near twin of the HARPS instrument in operation on the ESO 3.6m telescope 
at La Silla (Chile), covers the wavelength range from 3800\,\AA\ to 6900\,\AA\ with a 
resolving power of R$\sim$115,000. Each resolution element is sampled by 3.3 CCD pixels.

While out-of-transit data were obtained with the simultaneous Th-Ar calibration, 
the in-transit spectra were acquired with the second fibre on-sky to avoid 
the risk of contaminating the stellar spectrum by the calibration lamp.
Knowing the spectroscopic orbital parameters, in particular the semi-amplitude $K$ 
of the radial velocity (RV) curve and the systemic RV $\gamma$ is mandatory for a correct 
interpretation of the RM effect.
Following the successful acquisition of a spectral time series covering the Qatar-1\,b transit, 
we were prompted to gather additional HARPS-N data aiming to cover out-of-transit phases 
and improve the orbital solution. 

Thanks to the flexible scheduling of observations inside the GAPS programme, 
during following nights (September 5, 6, 7, 8, 9, and 11) we were able to obtain 
seven additional spectra evenly distributed over the different orbital phases.
The RV measurements of Qatar-1\, are reported in Table\,\ref{t:RV-data}.
All spectra were acquired with an exposure time of 900\,s.
The spectrograph is equipped with an exposure meter to accurately measure
the flux-weighted mean time of each exposure.
The RV measurements and corresponding errors were obtained using the HARPS-N on-line 
pipeline, based on the numerical cross-correlation function (CCF) method 
\citep{1996A&AS..119..373B} with the {\em weighted} and {\em cleaned}-mask modification 
\citep{2002A&A...388..632P}, by applying the K5 mask. 

To perform the detailed analysis of light curves and radial velocities data
the time-tag of each exposure was reduced to the solar system barycentric time using
the software Tempo2 \citep{2006MNRAS.369..655H} with the DE405 JPL Ephemerides
\citep{1998A&A...336..381S} in barycentric coordinate time (TCB) scale (SI units).
The assumed celestial coordinates of the source are:  
RA(J2000) = 20$^h$ 13$^m$ 31$^s$.615 DEC(J2000) = $+65^\circ$ 09$'$ 43$''$.48, 
with proper motions (in mas/yr) $\mu_\alpha=7.1$, $\mu_\delta=58.0$.
In Table\,\ref{t:RV-data}, we report the RV after barycentering with Tempo2 in TCB units.

\begin{table}
\caption{
HARPS-N RV measurements of Qatar-1.
Barycentric Julian Dates (BJD) in TCB units (SI units).
}
\label{t:RV-data}
\centering
\begin{tabular}{lcrrc}
\hline
\hline

 BJD (TCB)      &   RV      & error &   S/N$^\dagger$ & $\ddagger$\\ 
             & [\ms]    & [\ms] &                 &           \\
\hline
2456174.4128054 &  $-$38006.8 &  4.7  &  20             & b        \\
2456174.4247685 &  $-$38011.8 &  4.4  &  21             & b        \\
2456174.4355706 &  $-$38021.1 &  4.4  &  21             & i         \\
2456174.4463817 &  $-$38016.3 &  4.4  &  22             & i         \\
2456174.4572017 &  $-$38049.8 &  4.7  &  20             & i         \\
2456174.4680268 &  $-$38061.4 &  4.7  &  20             & i         \\
2456174.4788509 &  $-$38093.0 &  4.5  &  21             & i         \\
2456174.4896710 &  $-$38102.7 &  4.6  &  20             & i         \\
2456174.5005001 &  $-$38097.8 &  4.8  &  20             & i         \\
2456174.5113162 &  $-$38115.5 &  4.8  &  20             & b        \\
2456174.5232873 &  $-$38128.7 &  5.0  &  19             & b        \\
2456176.5702612 &  $-$38084.6 &  7.5  &  13             & o        \\
2456177.5788490 &  $-$38311.6 &  7.4  &  14             & o        \\
2456177.5954171 &  $-$38310.1 & 10.4  &  11             & o        \\
2456178.5619923 &  $-$37885.3 & 13.5  &   9             & o         \\
2456179.6162127 &  $-$37836.5 & 19.7  &   7             & o         \\
2456180.5669062 &  $-$38291.1 &  6.7  &  16             & o         \\
2456182.5996329 &  $-$37781.7 &  6.0  &  17             & o         \\
\hline
\multicolumn{5}{l}{
$\dagger$: signal-to-noise ratio per pixel at 5500 \AA; } \\
\multicolumn{5}{l}{$\ddagger$: i\,$\equiv$\,in-transit, o\,$\equiv$\,out-of-transit, b\,$\equiv$\,data taken just} \\
\multicolumn{5}{l}{before or after the transit, which are used for both the } \\
\multicolumn{5}{l}{orbital and the RM fit.} \\
\end{tabular}
\end{table}

\section{Ancillary data: transit photometry}
\label{sec:phot-obs}
New photometric data of Qatar-1\,b transits were obtained with the Asiago 1.82m 
and Calar Alto 1.23m telescopes. 
The journal of photometric observations is given in Table\,\ref{tab:phot_log},
while the photometric data are provided in Table\,\ref{tab:phot_data}.

\begin{table*}
\caption{Details of the photometric observations presented in this work. 
$N_{\mathrm{obs}}$ is the number of observations, Moon is the fractional 
illumination of the Moon at the midpoint of the transit, and $t_{\rm exp}$ is 
the exposure time in seconds. 
The aperture sizes are the radii in pixels of the software apertures for the star, 
inner sky, and outer sky, respectively. Scatter is the r.m.s.\ scatter of the data 
versus a fitted model in mmag. Times and dates are in UT.}
\label{tab:phot_log}
\centering %
\tiny %
\begin{tabular}{lcccrrccccc}
\hline\hline
Obs. & Camera/CCD & Date & UT(Start$\rightarrow$End) & $N_{\mathrm{obs}}$ & $t_{\rm exp}$ & Filter & Airmass & Moon & Aperture & Scatter \\
%                &                      &         &       UT(hh:mm)           &                                 & (s)                 &         &              &           &              & (mmag) \\
\hline %
Asiago & AFOSC/E2V 42-20        & 2011 05 29 & 22:$54\rightarrow02$:19 & 1218 &   7 & Cousins $R$ & $1.11\rightarrow1.06 $ & $7\% $ & 10,\,12,\,22 & 1.25\tablefootmark{a}  \\
CAHA  & SITE\#2b/2k$\times$2k  & 2011 08 25 & 23:$31\rightarrow04$:02 &  186 &  60 & Cousins $R$ & $1.14\rightarrow1.78 $ & $61\%$ & 12,\,30,\,45 & 1.65  \\
CAHA  & DLRMKIII/4k$\times$4k & 2012 07 21 & 20:$29\rightarrow00$:32 &  112 & 120 & Cousins $R$ & $1.45\rightarrow1.13 $ & $77\%$ & 10,\,30,\,50 & 0.78  \\
Asiago & AFOSC/E2V 42-20        & 2012 08 24 & 23:$09\rightarrow02$:13 & 1134 &   7 & Cousins $R$ & $1.09\rightarrow1.38 $ & $55\%$ &  5,\,6,\,16 & 1.41\tablefootmark{a}  \\
CAHA  & DLRMKIII/4k$\times$4k & 2012 09 10 & 00:$20\rightarrow03$:41 &   78 & 120 & Cousins $R$ & $1.30\rightarrow2.02 $ & $77\%$ & 21,\,33,\,50 & 0.87  \\
\hline %
\end{tabular}
\flushleft
\tablefoottext{a}{After binning}
\end{table*}

\subsection{Photometric observations from the Asiago Observatory}
\label{sec:Asiago}
Two complete transits of Qatar-1\,b were observed on 2011 May 29 and 2012 Aug 24 within the TASTE project 
\citep{2011A&A...527A..85N}.
The weather conditions in both nights were characterised by veils and thin cirrus, yet neither caused interruptions 
in the time series nor stops in the autoguide. 
Both observations were performed with the Asiago Faint Object Spectrograph and Camera (AFOSC) at the 
1.82m Copernico telescope in northern Italy. 
AFOSC is a classical focal-reducer camera equipped with a thinned, back-illuminated E2V 42-20 
$2048\times 2048$ CCD ($0.26''$/pix in unbinned mode),  providing a $9'\times 9'$ field of view. 
Both transits were observed in imaging mode through a standard Cousins $R$ filter, with a constant 
exposure time of 7 s. 
CCD windowing and $4\times 4$ binning were set to increase the time series duty-cycle, while a suitable set 
of reference stars was imaged on the same read-out window. 
Stellar images were defocused to $\sim 4''$ FWHM to avoid saturation and minimize flat-field residual errors. 

After a standard correction for bias and flat-field, the frames were reduced with the STARSKY code,
the TASTE photometric pipeline \citep{2013A&A...549A..30N}, which also includes a diagnostic tool to discard 
the data points most affected by transparency variations. 
Differential light curves were extracted by normalising the raw flux from the target with an optimally weighted 
average flux from the reference stars. 
Finally, light curves were corrected for systematic errors by decorrelating flux against external parameters 
(including e.g., star position on the detector, FWHM, background level, and airmass) and selecting the solution 
with the smallest off-transit scatter.

\subsection{Photometric observations from the Calar Alto Observatory}
\label{sec:CalarAlto}
Three transits of Qatar-1\,b were observed on 2011 August 25 and on 2012 July 21 and September 10, 
using the 1.23\,m telescope at the German-Spanish Calar Alto Observatory (CAHA) near Almer\'{i}a (Spain),
which was already successfully used to follow-up several transiting planets \citep{2012arXiv1212.3701M}.
During the 2011 observations, we used the 2k$\times$2k SITE\#2b optical CCD\footnote{This CCD 
has since been decomissioned.} with a FOV of $16^{\prime} \times 16^{\prime}$ and 
a pixel size of $24\,\mu$m, which translates to a pixel scale of $0.5^{\prime\prime}$ per pixel. 
We defocused the telescope in order to lower the flat-fielding noise.
The photometric data were gathered through a Cousins $R$ filter with an observing cadence of 60 sec.
To limit the dead time between exposures, we reduced the amount of time lost to CCD readout 
by reading out only a small window. Autoguiding was used. 
The two 2012 transits were obtained in filter Cousins $R$ with the new DLRMKIII camera, 
which is equipped with an E2V CCD231-84-NIMO-BI-DD sensor with 4k$\times$4k pixels 
of $15\mu$m and a FOV of $21^{\prime} \times 21^{\prime}$. Both transits were
observed with the CCD unbinned, the telescope heavily defocused, and an observing cadence of 120\,s.
The autoguider was properly focused to preserve a good pointing of the telescope.

The observations were analysed using the {\sc idl} pipeline from \citet{2009MNRAS.396.1023S}.
The images were debiased and flat-fielded using standard methods, then subjected to aperture 
photometry using the IDL task {\sc aper} \footnote{Part of the ``ASTROLIB subroutine library" 
distributed by NASA: idlastro.gsfc.nasa.gov/}.
Pointing variations were followed by cross-correlating each image against a reference image. 
We chose the aperture sizes and comparison stars that yielded the lowest scatter
in the final differential-photometry light curve. 
The relative weights of the comparison stars were optimised simultaneously by fitting a
second-order polynomial to the outside-transit observations to normalise them to unit flux.
% Table 3
\begin{table}
\caption{Excerpts of the R-band light curves (LC) of Qatar-1: this table will
be made available at the CDS. A portion is shown here for guidance 
regarding its form and content.} %
\label{tab:phot_data}   %{Table_2}
\centering %
\tiny
\begin{tabular}{llrr}
\hline\hline
Telescope/LC       & BJD (TCB)       & Diff. mag. & Uncertainty  \\
\hline %
Asiago \#1         & 2455711.4545895 & 0.0016630  & 0.0034607 \\
\vspace{0.1cm} %
Asiago \#1         & 2455711.4548095 & 0.0052700  & 0.0034568 \\
Asiago \#2         & 2456164.4647472 & 0.0025770  & 0.0027072 \\
\vspace{0.1cm} %
Asiago \#2         & 2456164.4648472 & 0.0021340  & 0.0027099 \\
CA 1.23m \#1       & 2455799.4805145 & $-$0.0002695 & 0.0014395 \\
\vspace{0.1cm} %
CA 1.23m \#1       & 2455799.4814715 & 0.0002191  & 0.0014412 \\
CA 1.23m \#2       & 2456130.3541082 & $-$0.0004088 & 0.0008100 \\
\vspace{0.1cm} %
CA 1.23m \#2       & 2456130.3561482 & 0.0007516  & 0.0007696 \\
CA 1.23m \#3       & 2456181.5142293 & $-$0.0000350 & 0.0005951 \\
\vspace{0.1cm} %
CA 1.23m \#3       & 2456181.5159694 & $-$0.0001722 & 0.0005738 \\
\hline %
\end{tabular}
\end{table}

\section{Data analysis}   % 
\label{Sec:analysis}

In this section we describe the analysis of the data presented in Sec.\,\ref{sec:phot-obs} and 
present the results we obtained.
The analysis of our five photometric data sets was performed by employing the software code 
JKTEBOP (version 28) to fit a transit light-curve (LC) model \citep{2008MNRAS.386.1644S}.

\subsection{New ephemerides for Qatar-1\,b transits } 

\begin{figure*}[]
    \begin{center}
\resizebox{1.0\hsize}{!}{\includegraphics[]{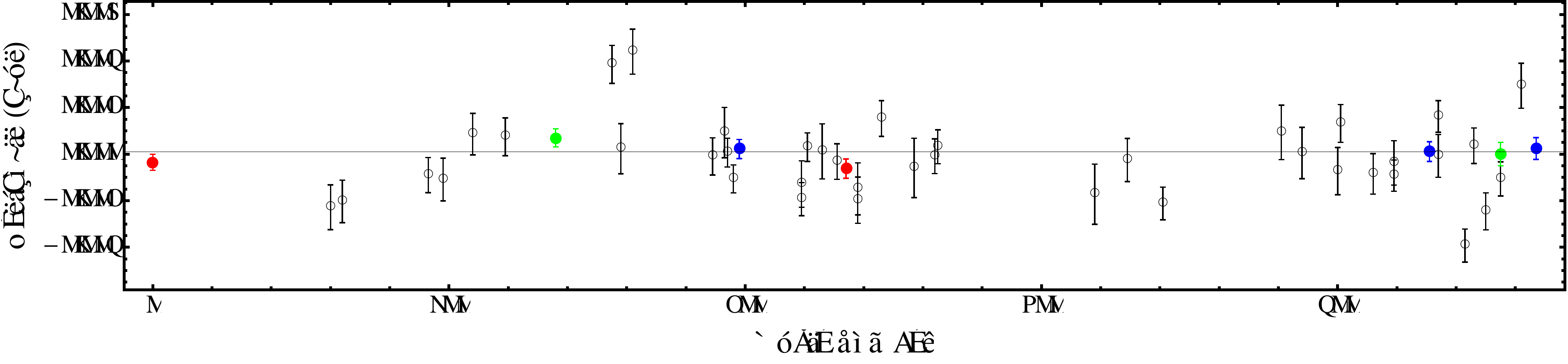}}
    \end{center}
\caption{ O$-$C diagramme for Qatar-1\, transit times. 
The small symbols in colour refer to the Asiago (green), Calar Alto (blue) and Alsubai et al. (2011) (red) data
used for redetermination of the ephemerides. Open circles represent transit times avaliable from the ETD.
}
\label{fig:transit_O-Cs}
\end{figure*}

As a first step we derived new ephemerides by combining the determinations of the mid-transit times 
from our five data sets with those derived from \citet{2011MNRAS.417..709A}.
All timings were placed on BJD(TCB) time system. 
The resulting measurements of transit midpoints were fitted with a straight line to obtain 
new orbital ephemerides:
\begin{equation} T_{0}=2\,455\,518.41094(28)+1.42002504(71) \times E , \end{equation}
where $E$ is the number of orbital cycles after the reference epoch (the midpoint of the first transit observed 
by \citealt{2011MNRAS.417..709A}) 
and the quantities in brackets denote the uncertainty in the last digits of the preceding number. 
The corresponding O$-$C diagramme is shown in Fig.\,\ref{fig:transit_O-Cs}, in which the mid-transit times 
available from the Exoplanet Transit Database (ETD)\footnote{http://var2.astro.cz/ETD/} are also displayed,
though they were not used in the fit.

\subsection{Combined solution of photometric light-curves }   
\label{sec:photsol}
The main parameters of the model are the fractional radii of the star and planet, r$_\star$ and r$_p$, 
defined as the stellar radius $R_\star$  and the planetary radius $R_\mathrm{p}$ scaled by the 
semi-major axis $a$, respectively, and the orbital inclination $i_p$. 

\begin{figure}[]
    \begin{center}
\resizebox{1.0\hsize}{!}{\includegraphics[]{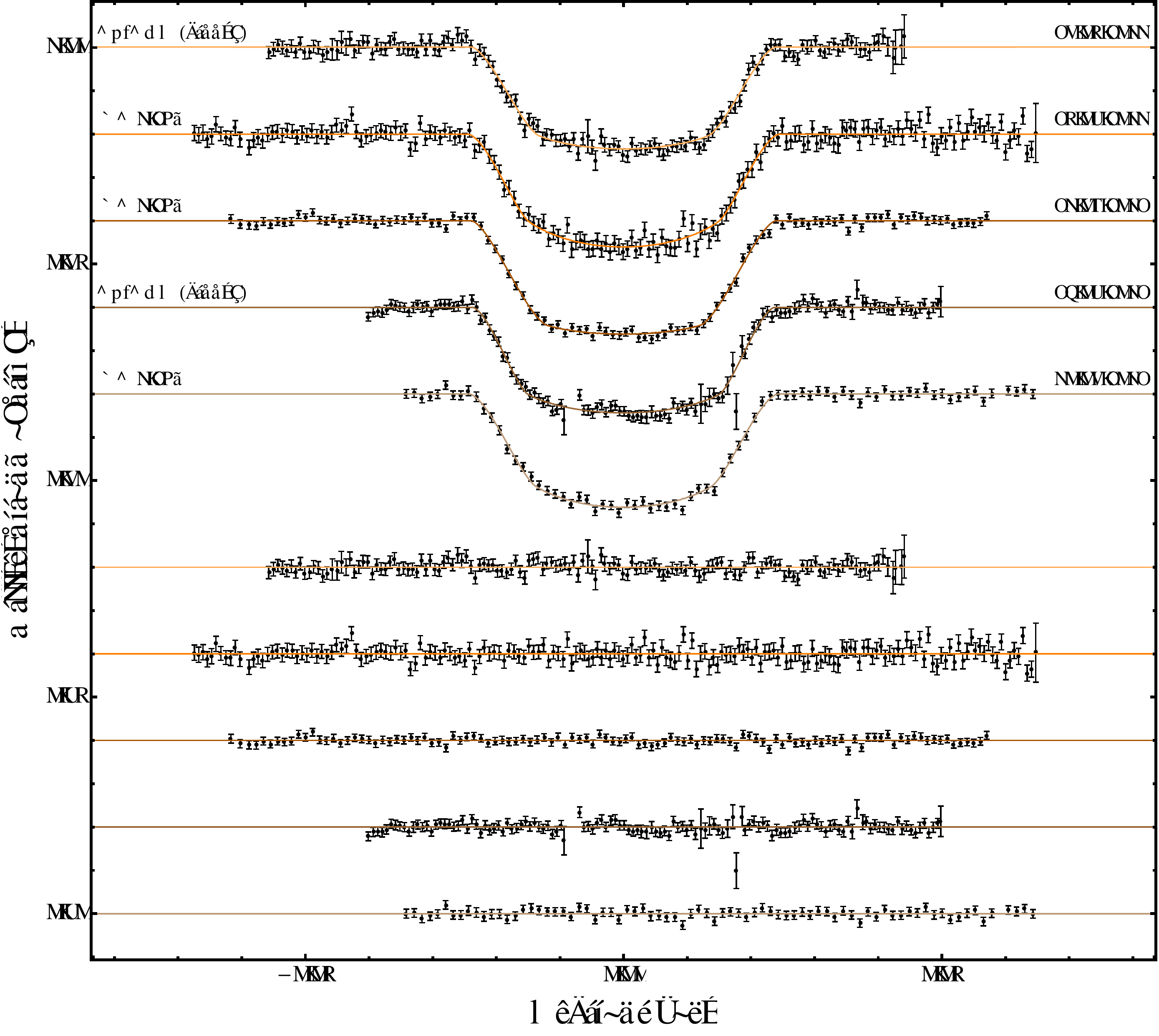}}
    \end{center}
\caption{Photometric transit light-curves of Qatar-1 and corresponding residuals
between data points and best-fit solution. }
\label{fig:transit_LCs}
\end{figure}

The LC solution was attempted using the linear and quadratic limb-darkening (LD) law, and with
the LD coefficients either fixed at the theoretical values tabulated by \citet{2011A&A...529A..75C}
or included as fit parameters.
We obtained  the best result by fitting the LD($u_1$) coefficient using a linear law.
We fitted individual light curves for the sum and ratio of the fractional radii $r_\mathrm{p}+r_\star$,
$r_\mathrm{p}/r_\star$, the orbital inclination $i$, the central transit time $T_0$, and the limb 
darkening coefficient LD($u_1$).
Evaluation of robust uncertainties on our best-fit parameters was performed by a bootstrap algorithm, 
i.e. generating for each light curve $10\,000$ resampled data sets to be fitted, and analysing
the resulting distribution \citep{2005MNRAS.363..529S}. 
The light-curves and their best-fitting models are shown in Fig.\,\ref{fig:transit_LCs}, whereas the results 
of our analysis are reported in Table\,\ref{tab:phot_CA} along with the weighted means (WMs) for each 
parameter evaluated over all five transits.
For some parameters ($r_\mathrm{p}/r_\star$ and LD($u_1$) the best-fit values are more scattered than
expected by applying Gaussian statistics.
A possible reason for these slight discrepancies could be intrinsic variations due to stellar activity of Qatar-1. 
Stellar spots, faculae and other active regions are known to alter the parameters inferred from 
photometric data, even when they are not occulted during the transit and thus no unusual feature 
is visible in the light curves \citep{2012A&A...539A.140B}. 
The ratio $r_\mathrm{p}/r_\star$ is one of the most critical parameters in this regard. 
We checked for long-term variability of Qatar-1 by performing differential photometry with the 
same comparison stars on the Asiago time series, A1 and A2, taken at the same site and airmass. 
We found a brightness variation between the two epochs of $+0.034 \pm 0.005$ and 
$+0.030 \pm 0.006$ mag (using two different reference stars), while the variation of our 
``check'' star is $-0.004 \pm 0.004$ mag. 
Although this finding is not fully conclusive, it suggests that stellar activity could play a role and 
possibly explain the differences in the LC solutions for different epochs. 

\begin{table*}[h]
\begin{center}
\small
\caption{Results from the analysis of individual and combined tranist light-curves}
\begin{tabular}{lcccclcc}     
\noalign{\smallskip}
\noalign{\smallskip}\hline
\noalign{\smallskip}\hline  \noalign{\smallskip}
Date/ID      &  r$_\star$+r$_p$ & r$_p$/ r$_\star$ &   $i_{\rm p}$     &    r$_\star$      &   r$_p$           & $b$ & LD ($u_1$) \\
\hline 
2011.05.29/A1 & 0.1850$\pm$0.0120 & 0.1441$\pm$0.0041 & 83.67$\pm$1.18 & 0.1617$\pm$0.0100 & 0.0233$\pm$0.0021 & 0.682$\pm$0.038 & 0.49$\pm$0.16 \\
2012.08.24/A2 & 0.1716$\pm$0.0093 & 0.1431$\pm$0.0022 & 84.99$\pm$0.74 & 0.1501$\pm$0.0077 & 0.0215$\pm$0.0014 & 0.582$\pm$0.047 & 0.50$\pm$0.09 \\
2011.08.25/C1 & 0.1793$\pm$0.0089 & 0.1465$\pm$0.0037 & 84.56$\pm$0.87 & 0.1564$\pm$0.0075 & 0.0229$\pm$0.0016 & 0.606$\pm$0.068 & 0.67$\pm$0.12 \\
2012.07.21/C2 & 0.1882$\pm$0.0043 & 0.1535$\pm$0.0009 & 83.36$\pm$0.35 & 0.1632$\pm$0.0036 & 0.0250$\pm$0.0006 & 0.709$\pm$0.022 & 0.37$\pm$0.11 \\
2012.09.10/C3 & 0.1834$\pm$0.0062 & 0.1496$\pm$0.0022 & 84.04$\pm$0.53 & 0.1595$\pm$0.0051 & 0.0239$\pm$0.0011 & 0.651$\pm$0.038 & 0.64$\pm$0.09 \\
\hline 
WM            & 0.1841$\pm$0.0030 & 0.1513$\pm$0.0008 & 83.82$\pm$0.25 & 0.1601$\pm$0.0025 & 0.0241$\pm$0.0005 & 0.675$\pm$0.016 & 0.54$\pm$0.05 \\
\hline 
\end{tabular}
\label{tab:phot_CA} \\
N.B. : r$_p$ = R$_p$/a; r$_\star$ = R$_\star$/a \\
\end{center}
\end{table*}

\subsection{Improved spectroscopic orbit solution }    
\label{sec:OrbSol}
The 11 RV measurements corresponding to out-of-transit phases (see Table\,\ref{t:RV-data}) 
form the data set we used to fit the spectroscopic orbit solution.

\begin{figure}[]
\begin{center}
\resizebox{1.0\hsize}{!}{\includegraphics[]{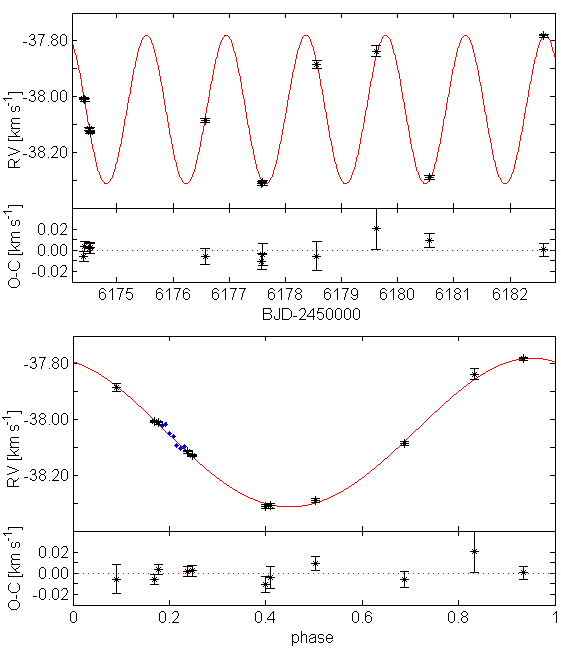}}
   \end{center}
 \caption{HARPS-N radial velocity data and best-fit spectroscopic orbit solution for Qatar-1. 
 Upper panel: RV curve versus BJD showing only the out-of-transit data 
 used for the orbital solution (solid red line), and the corresponding residuals. 
 Lower panel: phase-folded RV curve (solid red line) and data used for the orbital fit drawn 
 as black symbols. The RV measurements corresponding to in-transit phases are also displayed
 as light-blue symbols.
}
 \label{fig:RVs_orbit}
 \end{figure}

The solution of the RV curve was performed considering both the cases of circular and eccentric orbit.
The results of our RV curve fits are summarised in Table \ref{t:orb-par}, and the orbital solution 
corresponding to the fit obtained with the eccentricity as free parameter is shown in Fig.\,\ref{fig:RVs_orbit}.
The errors on the best-fit parameters were determined with the bootstrap method from 1000 mock data sets.

The comparison with the best-fit parameters reported in \citet{2011MNRAS.417..709A}
shows a number of significant differences. 
The Alsubai estimate of $\gamma$=-37835$\pm$63 \ms deviates from ours formally by more than 3$\sigma$. 
The most likely explanation for this discrepancy lies in a systematic RV zero-point difference between 
the two instruments used.
The higher precision of the HARPS-N RV measurements allows us to rule out an eccentric orbit for Qatar-1\,b.
Our determination of $e=0.020^{+0.011}_{-0.010}$ is compatible with a circular orbit within 2$\sigma$, 
as expected for close-in planets with orbital periods shorter than a few days \citep{2012MNRAS.422.3151H}.
Therefore, for the following analysis we adopted the best-fit parameters from the circular solution.
 
Finally, we obtained $K$=265.7 $\pm$  3.5 \ms, which is about 20\% higher than the value in 
\citet{2011MNRAS.417..709A}.
Accordingly, the estimates of the mass and density of Qatar-1\,b will be higher by the same percentage.

\begin{table}
\caption{Orbital parameters derived from the spectroscopic orbit solution of Qatar-1. 
          The parameters that were kept fixed in the fit are in italics. }
\label{t:orb-par}
\centering
\begin{tabular}{lcc}
\hline
\hline
                 \multicolumn{1}{c}{Orbital parameter} \\
\hline
                &  Eccentric fit               & Circular fit              \\ \cline{2-3}
$P$ [days]       &        \multicolumn{2}{c}{ {\it 1.42002504} } \\
$T_0$            &  ---                         & {\it 2\,456\,174.46251 } \\
$T_{peri}$ [BJD] & $2\,456\,174.17^{+0.13}_{-0.19}$ &   ---                     \\
$\gamma$ [\ms]   &  $-$38051.7 $\pm$ 2.5          & $-$38055.8 $\pm$  2.0       \\
$K_\ast$ [\ms]        &   266.8 $\pm$  3.4           &     265.7 $\pm$  3.5       \\
$e$              &  $0.020^{+0.011}_{-0.010}$   &    ---                    \\
$a_\ast$ [km]        & \multicolumn{2}{c}{ 5220 $\pm$ 82 } \\ 
$\omega$ [deg]   &  $17^{+33}_{-47}$          &    ---                    \\
$\chi^2$       &  1.184           &   1.380                 \\
\hline
\end{tabular}
\end{table}

 \begin{figure}[]
 \begin{center}
   \resizebox{1.0\hsize}{!}{\includegraphics[]{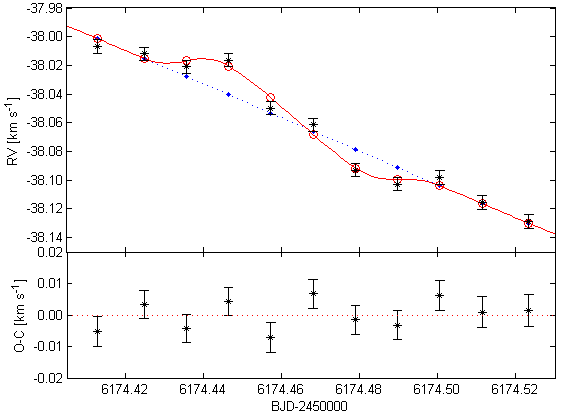}}
     \end{center}
 \caption{RM effect best-fit solution (solid, red line) with RV measurements (black star symbols) 
 and computed RV values (red open circles), yielding the O$-$C beneath. 
 The Keplerian radial velocity curve is also drawn as a blue dotted line. }
 \label{fig:RVs_RM}
 \end{figure}

\subsection{Analysis of the RM effect and determination of the spin-orbit alignment }
\label{sec:R-M}

To analyse the RM effect we implemented a numerical model based on the following assumptions.
We tried to reproduce the observed CCF by modelling the average photospheric line profile.
The stellar disc is sampled by a matrix of 2000$\times$2000 elements, each element being represented by 
a Gaussian line profile with a given width $\sigma_{el}$, Doppler-shifted according to the stellar rotation, 
and weighted by appropriate limb-darkening coefficients.   
The resulting line profile is then convolved by the instrumental profile (assumed to be Gaussian) of HARPS-N, 
$\sigma_{IP}=1.108$\,km\,s$^{-1}$. 
The model also takes into account the actual occulted area of the stellar photospheric disc and the smearing 
due to the planet's displacement during an exposure.
The corresponding RV shift is then computed by a Gaussian fit of this resulting line profile, 
analogously to the HARPS-N pipeline. The model depends on twelve parameters. 
The orbital period P$_{orb}$, mid-transit epoch $T_0$, transit duration $T_{14}$, star and planet radii
R$_{\star}$ and R$_{p}$, the linear limb-darkening coefficient LD$(u_1)$, and the impact parameter $b$ 
were fixed to the values we derived from the light-curve solution, while the star RV semi-amplitude $K$ was 
set to the best-fit value obtained from the spectroscopic orbit with null eccentricity.
The remaining parameters, i.e. the stellar projected rotational velocity v$\sin{I}$, the systemic velocity 
$\gamma$, the stellar disc resolution element line width $\sigma_{el}$, and the orbital obliquity 
$\lambda$ were treated as free parameters. 

By using a ``trust region" least-squares minimisation algorithm \citep{Branch1999,Byrd1988}, 
we obtained the following best-fit values: 
$\lambda = -8.4 \pm 7.1$\,deg, v$\sin{I} = 1.7 \pm 0.3 $\,km\,s$^{-1}$, 
$\sigma_{el} = 2238 \pm 155$\,m\,s$^{-1}$,  $\gamma = -38059.5 \pm 2.0$\,m\,s$^{-1}$,  
and normalised $\chi^2 = 1.217$.
The errors were determined with the same bootstrap method as for the orbital RV curve fit 
from 200 mock data sets.

\begin{figure*}[]
   \begin{center}
\resizebox{1.0\hsize}{!}{\includegraphics[]{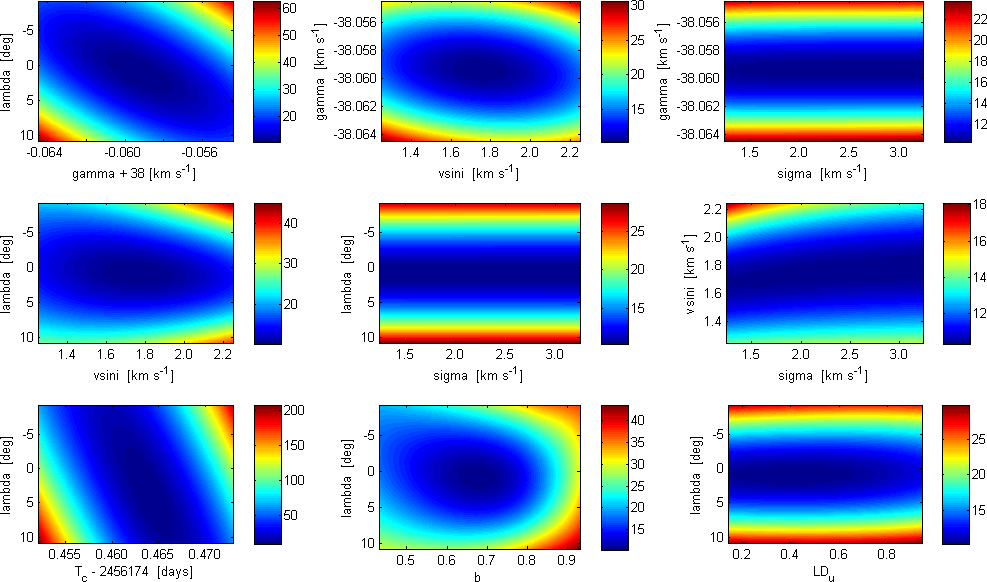}}
   \end{center}
\caption{$\chi^2$-maps showing the correlation between different couples of parameters.}
\label{fig:chi2-map}
\end{figure*}

Although an independent determination of the systemic velocity $\gamma$ is provided 
by the orbital RV curve fit, we preferred to treat $\gamma$ as a free parameter, 
because of its strong correlation with the parameter $\lambda$ (see Fig.\,\ref{fig:chi2-map}). 
In fact, by fixing the value of $\gamma$ to the value derived from the orbital RV curve fit, 
we obtained a best-fit value of $\lambda =-1^\circ$, with a normalised $\chi^2$ of 1.301.
By letting $\gamma$ free to vary, we also allowed for a possible RV shift caused by 
stellar spots, whose configuration on the stellar disc can be assumed 
to remain constant over the time-span of the transit.

This analysis shows that the system is well aligned.
The results are reported in Table\,\ref{t:all_par}.  
Fig.\,\ref{fig:chi2-map} shows the $\chi^2$-maps for various pairs of parameters.

A preliminary calibration of the FWHM of the CCF in terms of projected rotational velocity based on a sample 
of stars with planets observed as part of GAPS yields v$\sin{I}$=1.6$\pm$0.5\,km\,s$^{-1}$, consistent with the 
value derived as part of the RM effect modelling.

\section{Characterisation of the host star} 
\label{sec:host_char}
A proper determination of the stellar parameters is mandatory to obtain 
the physical properties of the planet.

Because of the uncertainty on the amount of interstellar extinction, photometric data alone cannot provide 
an accurate determination of the fundamental parameters of the star. 
According to the TASS Mark IV catalogue \citep{2006PASP..118.1666D}, the apparent V-band magnitude 
of Qatar-1 is V= 12.84\,mag. The K2\,V spectral type of the host star \citep[from ][]{2011MNRAS.417..709A} 
translates into an absolute magnitude of M$_{v}$=6.5\,mag \citep{1981Ap&SS..80..353S}. 
Depending on the amount of reddening, the distance to Qatar-1 is expected to span roughly between $\sim$190\,pc
for negligible extinction, and $\sim$130 pc for a normal interstellar extinction law 
(i.e., R$_{\rm v}=3.1$) and the maximum colour excess value (i.e., E$_{\rm B-V}=0.232\pm0.003$) obtained 
from the reddening map by \citet{1998ApJ...500..525S}. 

\subsection{Spectroscopic determination of stellar parameters}   
\label{sec:SpecSynt}
The HARPS-N spectra were used to perfom a spectroscopic characterisation of the 
host star Qatar-1 and estimate the effective temperature T$_{\rm eff}$, surface gravity 
$\log{g}$, projected rotational velocity v$\sin{I}$, and the iron abundance [Fe/H]).

The effective temperature was initially determined from the HARPS-N spectra by applying 
the method of equivalent width (EW) ratios of spectral absorption lines by means of the 
ARES\footnote{http://www.astro.up.pt/$\sim$sousasag/ares/} automatic code \citep{2007A&A...469..783S},
using the calibration for FGK dwarf stars by \citet{2010A&A...512A..13S}.
For this purpose, we used an average spectrum of Qatar-1 obtained by a weighted mean of the 
18 spectra available (after verifying that no contamination was present in the spectra taken with 
the simultaneous ThAr calibration), each properly shifted by the corresponding radial velocity to 
the rest wavelength frame, obtaining T$_{\rm eff}=4990\pm100$K.

\begin{figure}[]
   \begin{center}
\resizebox{1.0\hsize}{!}{\includegraphics[]{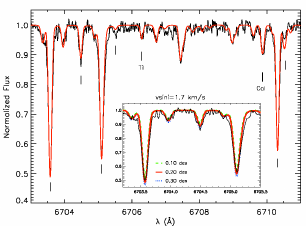}}
   \end{center}
\caption{Portion of Qatar-1 average spectrum (black solid line) and synthetic spectrum (red solid line) 
couplcorresponding to the following set of (fixed) parameters: $T_{\rm eff}=4910$\,K, $\log g=4.66$, 
$\xi=0.90$\,km\,s$^{-1}$, and $v \sin{I}=1.7$\,km\,s$^{-1}$. 
The positions of five iron lines and two lines of titanium and calcium are marked. 
The inset shows a zoom on the spectral window containing three Fe~{\sc i} lines, and the comparison with 
synthetic spectra corresponding to [\ion{Fe}{i}/H]=0.10 (green, long-dashed line), 0.20 (red, solid line), 
and 0.30 (blue, dotted line). }
\label{fig:spec-synth}
\end{figure}

Furthermore, the atmospheric stellar parameters were derived using the program MOOG 
\citep{1973ApJ...184..839S}, version 2010,  and through EW measurements of iron lines 
as described in detail by \citet{2012MNRAS.427.2905B}. 
In particular, the effective temperature was determined by imposing the condition that the \ion{Fe}{i} abundance 
does not depend on the excitation potential of the lines, the microturbulence velocity ($\xi$) was derived by imposing 
that the surface \ion{Fe}{i} abundance is independent on the line EWs, and the surface gravity was estimated by imposing 
the \ion{Fe}{i}/\ion{Fe}{ii} ionization equilibrium. Then, we were also able to measure the iron abundance of Qatar-1. 
This analysis was performed differentially with respect to the Sun. 
For this purpose, we analysed the Qatar-1 spectrum and a Ganymede spectrum acquired with HARPS at ESO
 \citep{2003Msngr.114...20M}. 
We thus obtained $\log n{(\ion{Fe}{i})_\odot}=\log n{(\ion{Fe}{ii})_\odot}=7.53\pm0.05$ for the Sun. 
In the end, we found the following atmospheric parameters and iron abundance for the star: 
$T_{\rm eff}=4820\pm100$\,K, $\log g=4.43\pm0.10$, $\xi=0.90\pm0.05$ km\,s$^{-1}$, 
[\ion{Fe}{i}/H]$=0.15\pm0.10$, and [\ion{Fe}{ii}/H]$=0.15\pm0.08$, which agree 
with the previous determinations by \citet{2011MNRAS.417..709A}.

In addition, the stellar parameters were derived independently using the methodology presented 
in \citet{2004A&A...415.1153S} and  \citet{2008A&A...487..373S}. 
This analysis was also based on iron line excitation and ionization equilibrium, using a grid of 
Kurucz (1997) model atmospheres and the 2002 version of MOOG. 
The ARES code was used to measure the equivalent widths for a sub-set of iron lines from the list in 
Sousa et al. (2008) that is especially suited for determining the parameters of stars with temperature 
below 5200\,K (for details see Tsantaki et al., A\&A, submitted).
The derived parameters are T$_\mathrm{eff}$=4786$\pm$95\,K, $\log{g}$=4.41$\pm$0.24, 
$\xi=0.78\pm0.29$ km\,s$^{-1}$, and [Fe/H]=0.18$\pm$0.06. 
These values agree fairly well with those reported above.

As a final consistency check, we exploited the fact that there is only one temperature for which the surface gravity 
from evolutionary models will meet the one obtained from the ionization equilibrium.
By imposing this condition, the surface gravity was obtained by comparison with the 
BaSTI\footnote{URL  http://wwwas.oats.inaf.it/IA2/BaSTI/} evolutionary models.
This yields the set of values $T_{\rm eff}=4910\pm100$\,K, $\log g=4.66\pm0.10$, and  [\ion{Fe}/H]$=0.20\pm0.10$.
Fig.\,\ref{fig:spec-synth} shows the observed and the synthetic spectrum corresponding to the latter set 
of parameters and adopting for the projected rotational velocity the value $v\sin{I}=1.7$km\,$s^{-1}$, 
as derived from the RM best-fit solution  (see Sect\,\ref{sec:R-M}). 
This set of values was finally adopted by us in the following.

 \subsection{SED, reddening, and distance}
\begin{figure}[]
   \begin{center}
\resizebox{1.05\hsize}{!}{\includegraphics[]{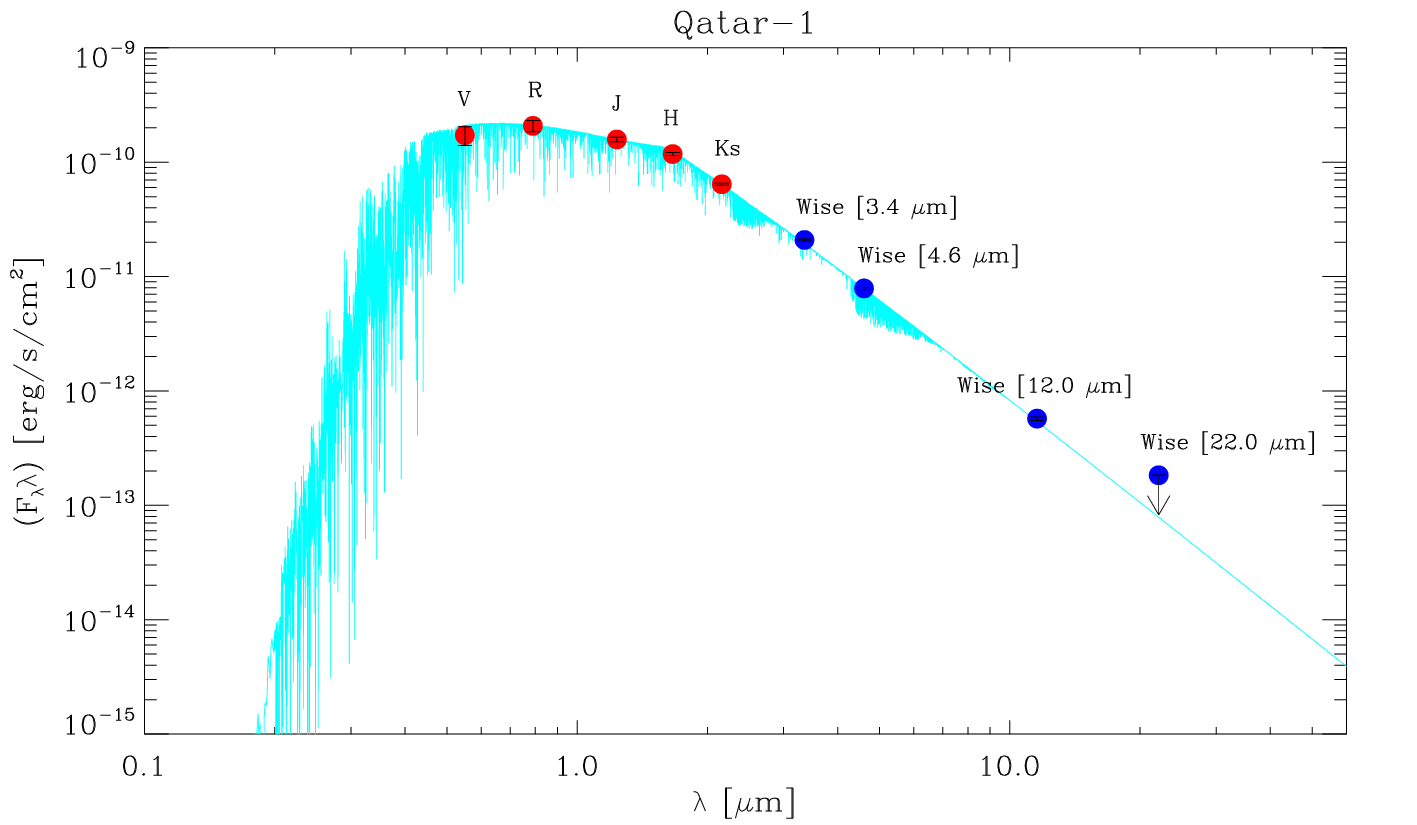}}
   \end{center}
\caption{De-reddened spectral energy distribution of Qatar-1, for $A_{\rm v}=0.1$ and d=195\,pc. 
Optical $V$ and $R$ photometric data are taken from the \emph{TASS Mark IV} catalogue \citep{2006PASP..118.1666D}. 
Infrared data are taken from the \emph{2MASS} and \emph{WISE} databases \citep{2003yCat.2246....0C,2012yCat.2311....0C}. 
The NextGen synthetic low-resolution spectrum  \citep{1999ApJ...512..377H} with the same photospheric 
parameters as Qatar-1 is over-plotted with a light-blue line.
% (see the online edition of the Journal for a colour version of this figure).
}
\label{fig:SED}
\end{figure}

We constructed the spectral energy distribution (SED) of Qatar-1 by merging the V and R optical magnitudes from 
the TASS Mark IV catalogue \citep{2006PASP..118.1666D} with the infrared photometry from the \emph{2MASS} 
and \emph{WISE} databases \citep{2003yCat.2246....0C,2012yCat.2311....0C}, as shown in Fig.\,\ref{fig:SED}.
We simultaneously fitted the colours encompassed by the SED to \emph{ad hoc} synthetic magnitudes 
derived from a \emph{NextGen} stellar atmosphere model \citep{1999ApJ...512..377H} with the same 
$T_{\rm eff}$, $log\,g$, and metallicity as the star. 
To estimate the interstellar extinction $A_{\rm v}$ and distance d to Qatar-1, we followed the method described in 
\citet{2008ApJ...687.1303G} and combined the availabe set of photometric data with the spectroscopically derived 
parameters.
Assuming a total-to-selective extinction $R_{\rm v}=A_{\rm v}/E_{\rm B-V}=3.1$  and a black body emission at the star 
effective temperature and radius yields an extinction $A_{\rm v}=0.10\pm0.10$\,mag and a distance to the star 
$d=195\pm25$\,pc, fairly consistent with the range of values estimated above.

\begin{figure}[]
   \begin{center}
\resizebox{1.0\hsize}{!}{\includegraphics[]{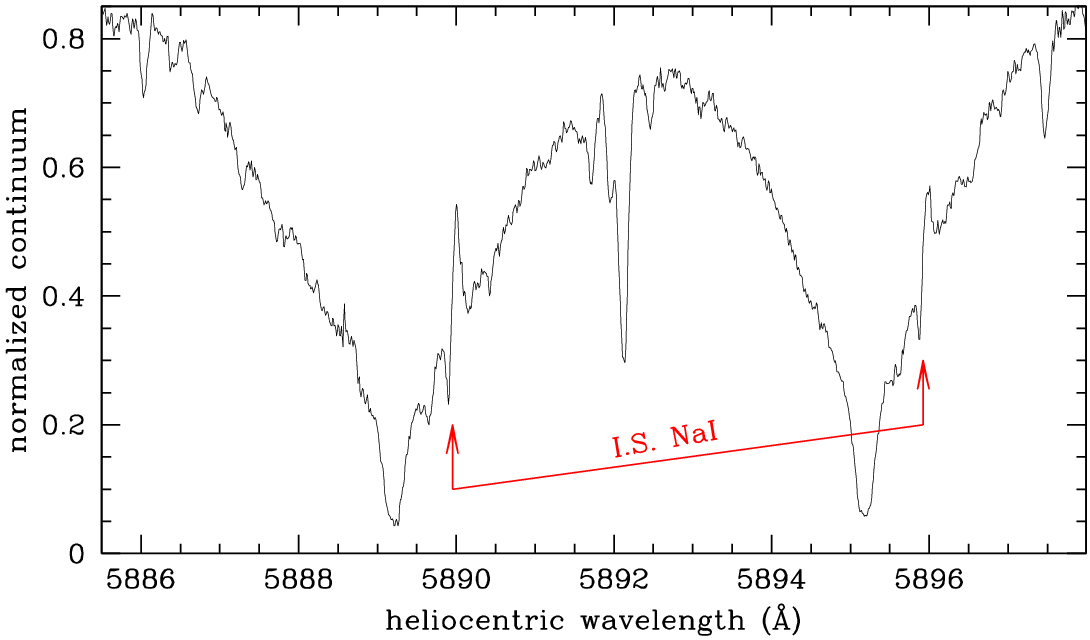}}
   \end{center}
\caption{Portion of the HARPS-N averaged spectrum of Qatar-1 highlighting the weak interstellar \ion{Na}{i} lines.}
\label{fig:NaD_IS}
\end{figure}

Weak interstellar components are visible within the broad and blended \ion{Na}{i} D$_{1,2}$ 
stellar doublet, as shown in Figure\,\ref{fig:NaD_IS}.  
Their equivalent widths are 0.017 and 0.009 ($\pm$0.001)\,\AA, respectively, i.e. very close 
to the asymptotic 2.0 limit ratio for optically thin 5889.951 and 5895.924\,\AA\ lines. 
Some contamination from the adjacent not perfectly subtracted night-sky lines is possible.
Adopting the \citet{1997A&A...318..269M} calibration between reddening and
equivalent width of interstellar \ion{Na}{i} 5889.951\,\AA\ line, the reddening
affecting Qatar-1 is $E_{\rm B-V}$=0.004 ($\pm$0.0005). 
 
Finally, using a method based on the near-infrared SED published in \citet{2006A&A...450..735M}, as described 
in Section 2.3 of \citet{2010ApJ...714..384R}, and adopting the values of $\log{g}$ and [Fe/H] derived 
from the spectroscopic analysis (though their impact on the final values is very small, similarly to the 
immunity of the infrared flux method to those parameters), we obtained an effective temperature of 
T$_\mathrm{eff}=4880\pm70$\,K.
The main source of error is the uncertainty in the extinction value, assumed to be Av=0.1+/-0.1 mag. 
Hence, the resulting effective temperature perfectly agrees with the spectroscopic determination. 

\subsection{Star and planet physiscal properties} 
Table\,\ref{t:all_par} summarises the physical properties we derived for the star and the planet in the Qatar-1 system.
To determine the stellar fundamental properties, we exploited quantities obtained directly from the analysis 
of the transit light curves following the prescriptions by \citet{2007ApJ...664.1190S}.
In particular, we derived the density of the star $\rho_\star=1.62\pm0.08\rho_\odot$ from the scaled stellar radius 
$R_\star/a$ and used it together with T$_\mathrm{eff}$ and [Fe/H] obtained from the spectroscopic anaslysis 
to infer the stellar mass and radius by comparison with stellar evolution models. 
Using the BaSTI models, this yields a stellar mass $M_\star=0.85\,M_\odot\pm0.03\,M_\odot$ and a radius 
$R_\star=0.80\pm0.05\,R_\odot$, which coincide with the values reported by  \citet{2011MNRAS.417..709A}.
In turn, the surface gravity of the planet, $\log g_{\rm p}=3.372\pm0.024$, and the planet equilibrium temperature,
$T_{\rm P}=1389 \pm 39$\,K, with their corresponding uncertainties were obtained following 
\citet{2007ApJ...664.1190S} and \citet{2011ApJ...729...54C}, respectively.
The planetary radius and mass are found to be $1.18\pm0.09$\,R$_{\rm J}$ and  $1.33\pm0.05$\,M$_{\rm J}$.
Hence, while only marginally larger in radius, the planet is found to be significantly more massive than
reported by \citet{2011MNRAS.417..709A}.

\section{Discussion}   % 
\label{Sec:results}
As discussed by \citet{2012A&A...540A..99E}, the radius of an exoplanet may be affected by a number of factors, 
which include the mass of the planet and its heavy element content as well as the irradiation received from the host star.
% The heating is expected to lead to larger planet radii, increased host-star metallicity to smaller planetary radii
% whereas no mass effect is expected on Jupiter-mass planets.
% 0.987 Rjup. Using this to estimate the Radius Anomaly I get 0.35. 
For Qatar-1\,b, we expect strong irradiation and tidal effects because of its proximity to the host star.
Below, we examine rotation and activity indicators and discuss the possible evolutionary effects 
caused by tidal interaction.

%%%%%%%%%%%%%%%%%   Star & Planet parameters  %%%%%%%%%%%%%%%%%
\begin{table*}
\caption[]{Summary of planet and star parameters for the Qatar-1 system. } 
\centering
\begin{centering}
\begin{tabular}{l c l c}
\hline\\
Parameter [Units]                                 & Symbol    &     Value   &  Method  \\
\hline\\
Transit epoch (BJD\_TCB$-$2450000.0)  [days]  & $T_0$      & $  5518.41094 \pm 0.00016 $  & (1) \\
Orbital period   [days]                           & $P$          & $ 1.42002504 \pm 0.00000071 $ & (1) \\
Transit duration  [days]                         & $T_{14}$  & $ 0.0678  \pm  0.0010 $   & (2)  \\ 
Planet/star area ratio     & $(R_{\rm p}/R_\star)^2$   & $  0.02280 \pm 0.00011 $   & (2)  \\
Planet/star radii ratio              & $R_{\rm p}/R_\star$   & $  0.1513 \pm 0.0008 $   & (2)  \\ 
Orbital inclination  [degrees]                  & $i_{\rm p}$          & $  83.82 \pm 0.25   $   & (2)  \\ 
Impact parameter                                 & $b$             & $ 0.675 \pm 0.016 $  & (2)  \\           
Scaled stellar radius                             & $R_\star/a$  & $  0.1601 \pm 0.0025 $   & (2)  \\           
Scaled planet radius                             & $R_p/a$      & $  0.0241 \pm 0.0005 $    & (2) \\           
Star reflex velocity  [km s$^{-1}$]        & $K_\star$    & $ 0.2657 \pm0.0035   $   & (3)  \\ 
Orbital eccentricity                              & $e$                & $ 0 $     & (3)   \\ 
Systemic velocity  [km s$^{-1}$]          & $\gamma$   & $  -38.0558 \pm0.0002 $   & (3) \\  

Star density  [$\rho_\odot$]                   & $\rho_\star$ & $  1.62 \pm 0.08   $    & (4) \\ 
Planet surface gravity   [cgs]                & $\log g_{\rm p}$   & $ 3.372 \pm 0.024 $   & (4)  \\           

Star mass  [M$_\odot$]                        & $M_\star$      & $  0.85 \pm 0.03  $    & (5)  \\  

Planet mass  [M$_{\rm J}$]                  & $M_{\rm p}$ & $   1.33 \pm 0.05    $   & (6)  \\    

Orbital semi-major axis  [AU]              &  $a$            & $ 0.02343\pm0.0012 $    & (7) \\  

Star radius   [R$_\odot$]                      & $R_\star$      & $  0.80 \pm 0.05  $   & (7)   \\ 
Star surface gravity                              & $\log g_\star$ & $  4.55 \pm 0.10  $   & (7)  \\
Planet radius [R$_{\rm J}$]                  & $R_{\rm p}$  & $   1.18 \pm 0.09  $   & (7)  \\ 
Planet density  [$\rho_{\rm J}$]            & $\rho_{\rm p}$  & $   0.80 \pm 0.20  $   & (7)  \\ 

Spin-orbit misalignment  [degrees]     & $\lambda$   &   $-8.4 \pm 7.1$           & (8)  \\ 
Star projected rotational velocity  [km s$^{-1}$] & v$\sin{I}$ &  $1.7 \pm 0.3$       & (8)      \\

Star effective temperature [K]               & $T_{\rm eff}$ & $ 4910 \pm 100  $    & (9)  \\  
% Star surface gravity                              & $\log g_\star$ & $  4.66 \pm 0.10  $   & (9)  \\
Stellar metallicity                                 & [Fe/H]            & $ 0.20 \pm 0.10$    & (9)   \\  
% Orbital eccentricity                              & $e$                & $ 0.02 \pm 0.01 $   \\ 
Star age  [Gyr]                                    & $Age_\star$       &  $ \approx 4.5 $    & (10)   \\ 
% Star gyrochronological age  [Gyr]      & $t_{gyro}$  & $  ...  $   & $ 1.5 \div 3.0 $  \\ 
% 
% 
Planet equilibrium temperature [K]       & $T_{\rm P}$      & $   1389 \pm 39 $    & (11)  \\
\hline\\

 \label{t:all_par}
\end{tabular}
\newline
{
(1) Derived by  combining the determinations of the mid-transit times from our photometric
data sets with those derived from \citet{2011MNRAS.417..709A}. Kept fixed in the 
orbital and Rossiter RV fitting.
(2) Derived as weighted mean of the best-fit determinations from our five transit light curves.
(3) Derived by fitting the orbital RV curve to out-of-transit RV measurements.
(4) Derived following \citet{2007ApJ...664.1190S}.
(5) Derived from evolutionary models.
(6) Derived by using the expression of the mass function.
(7) Derived from parameters above.
(8) Derived by fitting our RM effect model to in-transit RV measurements.
(9) Derived from spectroscopic analysis.
(10) Derived from gyrochronology.
(11) Derived following \citet{2011ApJ...729...54C}.
}
 
\end{centering}
\end{table*}

\subsection{Stellar rotation, activity, and age } 
\label{sec:rot_age}
Adopting the v$\sin{I}$ of $1.7\pm0.3$\,km\,s$^{-1}$ and radius of $0.80\pm0.05$\,R$_\odot$  for Qatar-1, 
and assuming that the star is viewed equator-on, the rotation period is calculated to be 
$24\pm6$\,\,days. The gyrochronological age, $t_{gyro}$, can hence be estimated using the relation given by 
\citet{2007ApJ...669.1167B} (equation 3). 
For a ${\rm (B-V)}=0.9$, corresponding to a $T_{\rm eff}$ of 4910\,K of the star, we obtain 
$t_{gyro}=1.7\pm1.1$\,Gyr.

On the other hand, adopting the revised gyrochronology by \citet{2010A&A...512A..77L} for stars with hot 
Jupiters, we estimate an age of $\sim4.5$\,Gyr that better agrees with that derived from stellar 
evolutionary tracks (cf. Fig. 5 of Alsubai et al. 2011).
 
Chromospheric emission is present in the \ion{Ca}{ii} H\&K lines. 
Unfortunately, the S/N of individual HARPS-N spectra in the wavelength range below 4000\AA\ is inadequate 
to obtain reliable measurements of the $\log R'_{HK}$ index. 
Therefore it is not possible to look for eventual variations in the chromospheric activity.
However, using the passband definitions in \citet{1991ApJS...76..383D} and a preliminary calibration of 
the flux in the \ion{Ca}{ii} H\&K lines, we obtained S$_{HK} = 0.389$ from the average spectrum used 
in Sec.\,\ref{sec:SpecSynt}. 
Then, by adopting ${\rm (B-V)}=0.9$ and the calibrations of \citet{1984ApJ...287..769N}, 
we find $\log{\rm R'_{HK}}=-4.60$. 
Therefore, Qatar-1 appears to be a moderately active star, similar to, e.g., HD\,189733 and TrES-3 
\citep{2010ApJ...720.1569K}.

Using the relations given by \citet{1984ApJ...287..769N} and \citet{2008ApJ...687.1264M}, the expected rotation 
period for $\log{\rm R'_{HK}}=-4.60$ and ${\rm (B-V)}=0.9$ would be 23.8\,d and 23.0\,d, respectively, 
hence within the range estimated from v$\sin{I}$.
However, using relation\,(3) from \citet{2008ApJ...687.1264M} of age as a function of $\log{\rm R'_{HK}}$, 
yields an age of about 1.1\,Gyr, which is considerably younger than the value inferred from gyrochronology
(see also \citealt{2013arXiv1301.5651P} for an updated analysis on the use of chromospheric activity as an 
age indicator).

Finally, we note that Qatar-1\,b fits well within the  $\log{\rm R'_{HK}}$-$\log g_{\rm p}$ correlation pointed out 
by \citet{2010ApJ...717L.138H} for close-in planets ($a< 0.1$\,AU) with $M_{\rm p}>0.1$\,$M_{\rm J}$, orbiting 
stars with 4200\,K$<T_{\rm eff} <6200$\,K.
It appears also to be consistent with the indication reported by \citet{2012A&A...540A..82K} that the level of 
the chromospheric activity of stars with $T_{\rm eff} <5500$\,K hosting close-in planets ($a<0.15$\,AU) may be 
enhanced by the presence of Jupiter-mass planets.

\subsection{Star--planet tidal interaction  }
The orbital period of the planet in the Qatar-1 system is much shorter than the rotation period 
of the star estimated in Sec.\,\ref{sec:rot_age}. 
Therefore, tides produce a decay of the orbit with a continuous transfer of angular momentum 
from the orbital motion to the spin of the star. The orbital angular momentum is only 0.28 
of the present stellar spin, which is insufficient to reach an equilibrium synchronous state. 
In other words, as a consequence of the tidal evolution, the planet is expected to be engulfed by the star. 
The timescale for the orbital decay $\tau_{\rm a}^{-1} \equiv (1/a)(da/dt)$, where $a$ is the orbital semi-major 
axis and $t$ the time, depends on the efficiency of the tidal dissipation inside the star that is parameterised 
by its modified tidal quality factor $Q^{\prime}_{\rm s}$. 
Considering the tidal dissipation efficiency required to account for orbital circularisation and synchronisation 
in close binary systems, \citet{2007ApJ...661.1180O} estimated that $Q^{\prime}_{\rm s}$ is of the order of $10^{6}$ 
for late-type main-sequence stars.  Adopting this value and the stellar and planetary parameters in 
Table\,\ref{t:all_par}, we obtain $\tau_{\rm a} \sim 0.5$\,Gyr for Qatar-1 using the tidal evolutionary model 
of \citet{2010A&A...516A..64L}.
The timescale for the alignment of the orbital angular momentum and the stellar spin can be estimated from 
the same tidal model and is $\tau_{\epsilon}^{-1} \equiv (1/\epsilon) (d\epsilon/dt)$, where $\epsilon$ is 
the obliquity of the stellar equator with respect to the orbital plane. 
Assuming  $Q^{\prime}_{\rm s} = 10^{6}$ and an initial obliquity $\epsilon (t=0) = 30^{\circ}$ yields 
$\tau_{\epsilon} \approx 0.2$~Gyr.

It is interesting to note that the timescale of orbital decay computed with $Q^{\prime}_{\rm s}=10^{6}$ is 
significantly shorter than the estimated age of the star from Sect\,\ref{sec:rot_age}. 
The difference is not as dramatic as in the case of, e.g., OGLE-TR-56 \citep{2007ApJ...661.1180O}, 
or Kepler-17 \citep{2012A&A...547A..37B}, where the orbital decay timescale turns out to be as short 
as 40-70 Myr, but this result suggests that the use of the same value of $Q^{\prime}_{\rm s}$ as derived for 
close stellar binary systems also for star-planet systems is not appropriate 
{\citep[cf. other analyses trying to constrain $Q_{\rm s}^{\prime}$, e.g., ][]{2002ApJ...568L.117P,2007ApJ...661.1180O,2007P&SS...55..643C,2009ApJ...698.1357J,2012A&A...545A...6P}. }
In principle, a lower limit on $Q^{\prime}$ can be established by an accurate timing of the transits 
extended over a baseline of several decades, which allows one to test the predictions of the above models. 
Specifically, for $Q^{\prime}_{\rm s} =10^{6}$, the orbital decay is expected to produce a variation 
of the observed epoch of mid-transit with respect to a constant-period ephemeris  of $O-C \sim 17$\,s in twenty years.

The tidal evolution of stellar obliquity has been recently discussed by \citet{2012MNRAS.423..486L}  
from a theoretical point of view, and by \citet{2012ApJ...757...18A} considering RM observation statistics.

\section{Summary and conclusions } 
\label{sec:conclusions}
We reported on observations of the RM effect for the Qatar-1 system 
and a new determination of the orbital solution based on HARPS-N at TNG observations. 
With these data we also obtained the spectroscopic characterisation of the host star.
Combining the radial velocity data with new transit photometry and with photometric data
from the literature, we derived new ephemerides and improved the orbital parameters for the system.
The most notable results from this analysis can be summarised as follows:
\begin{enumerate}
\item The new spectroscopic orbital solution is found to be consistent with a circular orbit and 
any significant eccentricity of the orbit can be ruled out. 
\item The RM effect was measured and the sky-projected obliquity was determined, from which 
we can conclude that the orbital plane of the system is well-aligned with the spin axis of the star.
This result and the derived properties of Qatar-1\,b appear to be in line with the general trend observed 
for close-in planets around stars cooler than about 6250\,K \citep{2012ApJ...757...18A}.
\item 
The host star is confirmed to be a slowly rotating, metal-rich K-dwarf, 
which yet is found to be moderately active, as inferred from the strength of the 
chromospheric emission in the \ion{Ca}{II} H\&K line cores and from changes in 
the photometric light-curves at different epochs, which hint at the presence of stellar spots.
\item The planet is found to be significantly more massive than previously estimated 
by \citet{2011MNRAS.417..709A}.
\item  
Qatar-1 appears to be consistent with the indication that the level of the chromospheric activity 
in stars cooler than 5500\,K that host close-in giant planets may be enhanced. 
An attempt at estimating the timescale for the orbital decay by  tidal dissipation was also 
presented, which deserves further investigation.

\end{enumerate}
 
\begin{acknowledgements}
{
The HARPS-N instrument has been built by the HARPS-N Consortium, a collaboration between 
the Geneva Observatory (PI Institute), the Harvard-Smithonian Center for Astrophysics, 
the University of St. Andrews, the University of Edinburgh, the Queen's University of Belfast, 
and INAF. This work has been partially supported by PRIN-INAF 2010.
This research made use of the SIMBAD database, operated at the CDS (Strasbourg, France). 
This work has made use of BaSTI web tools.
E.C., M.E. and K.B. thank Giulio Capasso, Fabrizio Cioffi and Andrea Di Dato for their support 
with the OAC computing facilities. E.C. thanks Paolo Molaro for stimulating discussions,
V.N. and G.P. acknowledge partial support by the Universit\'a di Padova through the 
``progetto di Ateneo" \#CPDA103591.
N.C.S. acknowledges the support by the European Research Council/European Community 
under the FP7 through Starting Grant agreement number 239953, as well as from 
Funda\c{c}\~ao para a Ci\^encia e a Tecnologia (FCT) through program Ci\^encia\,2007 
funded by FCT/MCTES (Portugal) and POPH/FSE (EC), and in the form of grants reference 
PTDC/CTE-AST/098528/2008 and PTDC/CTE-AST/120251/2010.
I.R. acknowledges financial support from the Spanish Ministry of Economy 
and Competitiveness (MINECO) and the ''Fondo Europeo de Desarrollo 
Regional" (FEDER) through grants AYA2009-06934 and AYA2012-39612-C03-01.
A.T. is supported by the Swiss National Science Foundation fellowship number PBGEP2\_145594.
}
\end{acknowledgements}

\bibliography{ms.bbl}
\bibliographystyle{aa}

\end{document}